\pdfoutput=1 
\documentclass[doublespacing]{elsart}
\usepackage{icarus}

\usepackage{amssymb, amsmath}
\usepackage{booktabs} 

\usepackage[final]{graphicx} 

 \usepackage[squaren, textstyle]{SIunits}
 \newcommand{\arcsec}{\arcsecond} 
 
 \newcommand{\micron}{\micro\metre}

\usepackage{natbib}

\usepackage[breaklinks,bookmarksnumbered,pdftex,
pdftitle={Eclipsing Binary Trojan Asteroid Patroclus: Thermal Inertia From Spitzer Observations},
pdfauthor={Michael Mueller et al.},
pdfkeywords={Asteroids, Thermal inertia, Size, Albedo, Spitzer observations}
]{hyperref} 

\newcommand{\pV}{\ensuremath{p_\text{V}{}}}
\newcommand{\pv}{\pV}
\newcommand{\AU}{\ensuremath{\text{AU}}}

\newcommand{\TIunit}{\joule\usk\power{\second}{-1/2}\power{\kelvin}{-1}\power{\metre}{-2}}

\newcommand{\gramm}{\gram}
\newcommand{\cm}{\centi\metre}
\newcommand{\km}{\kilo\metre}
\newcommand{\kg}{\kilo\gram}
\newcommand{\hr}{\hour}
\newcommand{\Area}{\ensuremath{\mathcal{A}}}

 \newcommand{\chitwo}{\ensuremath{\chi^2{}}}

\newcommand{\water}{\ensuremath{\text{H}_2\text{O}}}

\newcommand{\AandA} {Astron.\ Astrophys.}
\newcommand{\aanda}{\AandA}

\newcommand{\AJ}{Astron.\ J.}

\newcommand{\ApJ}  {Astrophys.\ J.}
\newcommand{\ApJS} {\ApJ\ Suppl.\ Ser.}

\newcommand{\baas}{Bull.\ AAS}

\newcommand{\Icarus} {Icarus}
\newcommand{\JGR}    {J.\ Geophys.\ Res.}

\newcommand{\MNofRAS}{Mon.\ Not.\ R.\ Ast.\ Soc.}
\newcommand{\mnras}{\MNofRAS}

\newcommand{\Nature} {Nature}

\newcommand{\pss}{Plan.\ Space Sci.}
\newcommand{\PSS}{\pss}
 
\newcommand{\Science}{Science}

\newcommand{\journalref}[6]{#1\ #2. \textit{#3.} #4 \textbf{#5}, #6.}

\newcommand{\journalinpressref}[4]{#1\ #2. \textit{#3.} #4, \textit{in press.}}

\newcommand{\bookchapterref}[5]{#1\ #2. \textit{#3.} #5, pp.\ #4.}

\newcommand{\bookforchapter}[4]{In: #1 \textit{#2.} #3, #4}

\newcommand{\editors}[1]{#1\ (Eds.)}

\newcommand{\AsteroidsII}{\bookforchapter{\editors{Binzel, R.P., Gehrels, T., Matthews, M.S.}}{Asteroids II}{Univ.\ of Arizona Press}{Tucson}}
\newcommand{\AsteroidsIII}{\bookforchapter{\editors{Bottke, W.F., Paolicchi, P., Binzel, R.P., Cellino, A.}}{Asteroids III}{Univ.\ of Arizona Press}{Tucson}}

\newcommand{\ACMV}{\bookforchapter{\editors{Lazzaro, D., Ferraz-Mello, S., Fern\'andez, J.A.}}{Asteroids, Comets, and Meteors 2005}{Cambridge University Press}{Cambridge, UK}}
\newcommand{\Jupiterbook}{\bookforchapter{\editors{Bagenal, F., Dowling, T.E., McKinnon, W.B.}}{Jupiter: The planet, satellites, and magnetosphere}{Cambridge University Press}{Cambridge, UK}}
\newcommand{\KBObook}{\bookforchapter{\editors{Barucci, M.A., Boehnhardt, H., Cruikshank, D.P., Morbidelli, A.}}{The Solar System beyond Neptune}{Univ.\ of Arizona Press}{Tucson}}

\newcommand{\Delbo}{Delbo'}

\newcommand{\Fernandez}{Fern\'{a}ndez} 
\newcommand{\Ivezic}{Ivezi\'c}
\newcommand{\Juric}{Juri\'c}

\newcommand{\Mueller}{M\"uller}
\newcommand{\Nesvorny}{Nesvorn\'{y}}

\newcommand{\Szabo}{Szab\'o}

\renewcommand{\eqref}[1]{eqn.\ \ref{#1}}
\newcommand{\Eqref}[1]{Eqn.\ \ref{#1}}
\newcommand{\tableref}[1]{table \ref{#1}}
\newcommand{\Tableref}[1]{Table \ref{#1}}
\newcommand{\figref}[1]{\Figref{#1}}
\newcommand{\Figref}[1]{Fig.\ \ref{#1}}
\newcommand{\sectref}[1]{sect.\ \ref{#1}}

\newcommand{\seesect}[1]{(see \sectref{#1})}

\begin{document}

\begin{frontmatter}

\title{Eclipsing Binary Trojan Asteroid Patroclus: Thermal Inertia from Spitzer Observations}
\author[ua,dlr]{Michael Mueller\corauthref{cor}},
\author[seti,Berkeley,IMCCE]{Franck Marchis},
\author[Tennessee]{Joshua P.\ Emery},
\author[dlr]{Alan W.\ Harris},
\author[dlr]{Stefano Mottola}
\author[IMCCE]{Daniel Hestroffer},
\author[IMCCE]{J\'erome Berthier}
\author[oat]{Mario di Martino}

\address[ua]{University of Arizona, Steward Observatory, 933 N Cherry Ave, Tucson AZ 85721, USA}
\address[dlr]{DLR Institute of Planetary Research, Rutherfordstr.\ 2, 12489 Berlin, Germany}
\address[seti]{SETI Institute, 515 N Whisman Road, Mountain View CA 94043, USA}
\address[Berkeley]{University of California at Berkeley, Department of Astronomy, 601 Campbell Hall, Berkeley, CA 94720, USA}
\address[IMCCE]{Observatoire de Paris, Institut de M\'ecanique C\'eleste et de Calcul des \'Eph\'em\'erides, UMR8028 CNRS, 77 av.\ Denfert-Rochereau, 75014 Paris, France}
\address[Tennessee]{University of Tennessee, 306 Earth and Planetary Sciences Bldg, 1412 Circle Drive, Knoxville TN 37996, USA}
\address[oat]{INAF, Osservatorio Astronomico di Torino, Via Osservatorio 20, Pino Torinese, IT 10025 Torino, Italy}

\corauth[cor]{Corresponding author: michael.mueller@oca.eu (now at Observatoire de la C\^ote d'Azur)}

\date{2009 March 2, revised 2009 Jul 8}

\end{frontmatter}

\begin{flushleft}
\vspace{1cm}
Number of pages: \pageref{lastpage} \\
Number of tables: \ref{lasttable}\\
Number of figures: \ref{lastfig}\\
\end{flushleft}

\begin{pagetwo}{Thermal Inertia of Trojan Binary Patroclus}

Michael Mueller \\
Steward Observatory \\
933 N Cherry Ave \\
Tucson AZ 85721 \\
USA \\ 
\\
Email: michael.mueller@oca.eu \\

\end{pagetwo}

\begin{abstract}

We present mid-infrared (8--\unit{33}{\micron}) observations of the binary L5-Trojan system (617) Patroclus-Menoetius  before, during, and after two shadowing events, using the Infrared Spectrograph (IRS) on board the Spitzer Space Telescope.
For the first time, we effectively observe changes in asteroid surface temperature in real time, allowing the thermal inertia to be determined very directly.
A new detailed binary thermophysical model is presented which accounts for the system's known mutual orbit, arbitrary component shapes, and 
thermal conduction in the presence of eclipses.

We
obtain two local thermal-inertia values, representative of the respective shadowed areas:
$21\pm\unit{14}{\TIunit}$ and $6.4\pm\unit{1.6}{\TIunit}$.
The average thermal inertia is estimated to be $20\pm\unit{15}{\TIunit}$, potentially with significant surface heterogeneity.
This first thermal-inertia measurement for a Trojan asteroid indicates
 a surface covered in  fine regolith.
Independently, we establish the presence of fine-grained ($<$ a few \micron) silicates on the surface, based on emissivity features near 10 and \unit{20}{\micron} similar to those previously found on other Trojans.

We also report  $V$-band observations and report a lightcurve with complete rotational coverage. 
The lightcurve has a low amplitude of $0.070\pm\unit{0.005}{\text{mag}}$ peak-to-peak, implying a roughly spherical shape for both components,
and is single-periodic with a period
($103.02 \pm \unit{0.40}{\hr}$) equal to the period of the mutual orbit, 
indicating that the system is fully synchronized.

The diameters of Patroclus and Menoetius are 
$106\pm11$ and $98\pm\unit{10}{\km}$, respectively, in agreement with previous findings.
Taken together with the system's known total mass, this implies a  bulk mass density of $1.08\pm\unit{0.33}{\gramm\usk\power{\cm}{-3}}$, significantly below the mass density of L4-Trojan asteroid (624) Hektor and suggesting a bulk composition dominated by water ice.

All known physical properties of Patroclus, arguably the best studied Trojan asteroid, are consistent with those expected in  icy objects with devolatilized surface (extinct comets), consistent with what might be implied by recent dynamical modeling in the framework of the Nice Model.

\end{abstract}


\section{Introduction}
\label{sect:intro}

\subsection{Thermal inertia}
\label{sect:intro:TI}

Thermal inertia is a measure of the resistance to changes in surface temperature. 
A hypothetical zero-thermal-inertia asteroid would be in instantaneous thermal equilibrium with insolation
and display a prominent sub-solar temperature peak.
In general,  thermal inertia 
causes the surface temperature distribution to have reduced contrast and be asymmetric with respect to the sub-solar point, with the maximum shifted to the afternoon side.
The influence of thermal inertia on the temperature distribution is maximum when the spin vector is perpendicular to the solar direction and zero if it points towards the Sun.

 Thermal inertia is defined as $\Gamma=\sqrt{\kappa\rho c}$, with thermal conductivity $\kappa$, surface bulk mass density $\rho$, and heat capacity $c$.
The range in which $\rho$ and $c$ can plausibly vary is much more restricted than for $\kappa$, which can vary by several orders of magnitude \citep[see, e.g.,][Sect.\  3.2.2.b for a detailed discussion]{Mueller2007}.

 Examples of low and high thermal inertia cases would be a dusty, thermally insulating surface such as 
that of the Moon, and a rocky, thermally conductive surface, respectively.
It is qualitatively clear that for fine-grained regolith (where the typical grain size is not large compared to the penetration depth of the heat wave), thermal conductivity and  thermal inertia decrease with decreasing grain size.
This is confirmed by laboratory studies
under a simulated Martian atmosphere \citep{Presley1997}, but no quantitative data are available for vacuum conditions.

The thermal inertia of the largest few main-belt asteroids is known to be low and indicative of a surface covered in fine regolith \citep[e.g.][]{Spencer1990,MuellerLagerros1998}.
It was recently found that the thermal inertia of 
$D<\unit{100}{\km}$ main-belt asteroids and  near-Earth asteroids
is increased relative to large main-belt asteroids, more so for smaller objects, but that it stays significantly below values expected for bare-rock surfaces \citep{Mueller2007,Delbo2007,Delbo2009}.
No quantitative measurements of the thermal inertia of Trojans have been published.

The thermal inertia of asteroids is typically
determined indirectly, 
from multi-filter observations 
in the challenging mid-infrared wavelength range 
at widely-spaced phase angles, such that  the sub-observer point is located at different rotational phases (``local times'') and information on the diurnal temperature distribution can be obtained.  
Depending on the apparent target motion, this may require observations stretched out over a significant time span.

Eclipse-induced shadowing, on the other hand, offers an opportunity to determine the thermal inertia  more directly, by monitoring the shadowing-induced cooling and the later warming up.
This technique 
was pioneered by \citet{PettitNicholson1930}, who
determined the thermal inertia of the Moon for the first time from thermal observations during a total lunar eclipse.
\citet{MorrisonCruikshank1973} report observations of the Galilean satellites as they were eclipsed by Jupiter; \citet{Neugebauer2005} report observations of  Iapetus during eclipses by Saturn's rings; eclipse observations of Saturn's satellites using the Cassini spacecraft are ongoing \citep{Pearl2008}.

Asteroids are not frequently shadowed by planets.
Binary asteroids, however, can undergo mutual eclipses, during which one component shadows the other. 
Recent progress now allows mutual eclipse events in an increasing number of binary asteroid systems to be reliably predicted.

\subsection{Binary Trojan asteroid Patroclus}
\label{sect:intro:Patroclus}

Our target, (617) Patroclus, is one of the few known binary systems in the population of  Trojan asteroids, which are in 1:1 orbital resonance with Jupiter, close to the Lagrangian points L4 and L5.
Trojan orbits are stable over most of the age of the Solar System \citep{Levison1997}.
Their origin is currently under debate:
While they were long believed to have formed near their present position \citep[see, e.g.,][]{Marzari2002}, \citet{Morbidelli2005,Morbidelli2009} argue in the context of the 'Nice model' that Trojans were captured
by Jupiter
during a chaotic phase in the early Solar System, which was caused by a mean-motion resonance between Jupiter and Saturn.
In this model, Trojans share 
a volatile-rich parent population with comets and Trans-Neptunian Objects
but have spent parts of their lifetime at relatively small heliocentric distances (potentially enough to leave their surface devolatilized) before being trapped in their current orbits.

Trojans have generally low albedos of $\pv\sim0.04$ and virtually featureless, highly reddened reflection spectra in the visible and near-IR wavelength ranges; in both respects, they resemble cometary nuclei 
\citep[see][for a recent review]{Dotto2008}.
Mid-IR spectra of three Trojans have revealed the presence of fine-grained silicates on the surfaces \citep{Emery2006}.
Trojans smaller than some \unit{70}{\kilo\metre} in diameter appear to be collisional fragments, while larger bodies  appear to be primordial ``accretion survivors,'' i.e.\ bodies whose current form and internal structure have remained unchanged since the time of their formation \citep{Binzel1992,Jewitt2000}.

(617) Patroclus was discovered in October 1906 by August Kopff. 
Following (588) Achilles, which had been discovered in February 1906, Patroclus was the second known Trojan and the first known object in the L5 swarm.
With an absolute magnitude in the HG system 
\citep[V magnitude normalized to a standard geometry; see][]{HG} 
of $H=8.19$, Patroclus is among the largest Trojans.
Based on IRAS observations, \citet{SIMPS} report a diameter of 
$D=140.9\pm\unit{4.7}{\km}$, corresponding to a geometric albedo of $\pv=0.047\pm0.003$
(see \eqref{eq:dpvh}  below).
\citet{Fernandez2003} obtained Keck \unit{12.5}{\micron} observations of Patroclus and derived a diameter of $166.0\pm\unit{4.8}{\km}$ ($\pv=0.034\pm0.004$). 
Note that the quoted diameters are area-equivalent diameters $D_\Area$ of the system as a whole. $D_\Area$ is related to the components' diameters, $D_1$ and $D_2$, through
$D_\Area{}^2 = D_1{}^2 + D_2{}^2$.

Patroclus was found to be binary by \citet{Merline2001}.
The system's mutual orbit  was first determined by \citet{Marchis2006} based on spatially resolved adaptive-optics observations.
Subsequently the secondary component was christened Menoetius.
\citet{PatroNew} provide an updated orbit model (consistent with that by \citeauthor{Marchis2006}) based on additional 
spatially non-resolved optical photometry measurements during mutual events.
The \citeauthor{PatroNew}\ model is used throughout this paper.
The system's mutual orbit is purely Keplerian (no precession) and circular 
 (eccentricity $\leq 0.001$).
The center-to-center separation of the two components is $654\pm\unit{36}{\km}$,  corresponding to a maximum angular separation around \unit{0.2}{\arcsecond}.
The J2000 ecliptic coordinates of the spin-pole orientation are
 $\lambda = 241.37\pm\unit{0.33}{\degree}$, $\beta=-57.35\pm\unit{0.36}{\degree}$, the orbital period equals $102.94\pm\unit{0.11}{\hour}$. 
This orbit model reproduces the adaptive-optics observations and the morphology of eclipse-induced lightcurves very well. Predicted and measured eclipse timing  may differ by up to a few hours (= a few percent of the orbital period). Work is under way to resolve these slight timing differences. No discernible change in the model's spin-axis orientation is expected to result from this.

\citet{Marchis2006} report a component-to-component magnitude difference of $\sim\unit{0.17}{\text{mag}}$; the same magnitude difference was found for two filters centered at  1.6 and \unit{2.2}{\micron}, respectively, which \citeauthor{Marchis2006}\ take as an indication of similar surface composition.
Assuming identical albedo, the corresponding diameter ratio is $D_1/D_2 \sim 1.082$.

Through Newton's parametrization of Kepler's Third Law, the orbital parameters imply a total system mass of $(1.20\pm0.20)\times \unit{$10^{18}$}{\kg}$.
\citeauthor{Marchis2006}\ used the size estimate by \citet{Fernandez2003}\ to 
infer a bulk mass density of only \unit{$0.8^{+0.2}_{-0.1}$}{\gramm\usk\power{\cm}{-3}}, 
compatible with a composition dominated by water ice and moderate porosity.
\citeauthor{PatroNew}\ determined the components' diameters independently, from eclipse timing, to be $D_1=112\pm\unit{16}{\km}$ and $D_2=103\pm\unit{15}{\km}$, and a bulk mass density of \unit{$0.90^{+0.19}_{-0.33}$}{\gramm\usk\power{\cm}{-3}}.

\subsection{Work presented in this paper}
\label{sect:intro:project}

The \citet{Marchis2006} orbit model allowed a series of 
mutual events  to be predicted,
including two shadowing events in June 2006 which we observed using the Spitzer Space Telescope.
We observed the system's thermal emission before, during, and after the events \seesect{sect:obs}.
Although the system was not spatially resolved, Spitzer's high sensitivity allowed us to 
effectively observe the eclipse-induced cooling and later warming up of shadowed surface elements in real time.
\emph{No such observations have been performed before. 
We present the first \emph{direct} measurement of the thermal inertia of an asteroid.}
A new thermophysical model for eclipsing binary asteroids has been developed to facilitate the data analysis \seesect{sect:TPM}.

We also obtained Patroclus' mid-IR emissivity spectrum from our Spitzer data \seesect{sect:features}, and its visible lightcurve from  ground-based observations \seesect{sect:visible}.

\section{Visible lightcurve observations}
\label{sect:visible}

\begin{table}
\caption{Optical observations: 
Observation mid time, observer-centric position ($\lambda$ and $\beta$ in J2000 ecliptic coordinates), solar phase angle $\alpha$, heliocentric distance $r$, and geocentric distance $\Delta$. $V(1,\alpha)$ denotes the observed mean magnitude corrected for heliocentric and geocentric distance, but not for phase angle.}
\label{table:visible}
\begin{tabular}{r|cccccc}
\toprule
UT date & $\lambda$ & $\beta$ & $\alpha$ & $r$ & $\Delta$ & $V(1,\alpha)$ \\
(Apr 1996) & (deg) & (deg) & (deg) & (AU) & (AU) & \\
\midrule
9.2 & 212.5 & 6.8 & 2.4 & 5.7592 & 4.7839 & 8.49 \\
10.2 & 212.4 & 6.8 & 2.3 & 5.7585 & 4.7795 & 8.49 \\
11.3 & 212.3 & 6.7 & 2.2 & 5.7578 & 4.7795 & 8.49 \\
12.3 & 212.1 & 6.7 & 2.0 & 5.7572 & 4.7715 & 8.48 \\
13.2 & 212.0 & 6.7 & 1.9 & 5.7566 & 4.7682 & 8.45 \\
15.2 & 211.8 & 6.6 & 1.6 & 5.7553 & 4.7621 & 8.44 \\
16.2 & 211.6 & 6.6 & 1.5 & 5.7546 & 4.7596 & 8.44 \\
17.3 & 211.5 & 6.6 & 1.3 & 5.7539 & 4.7573 & 8.42 \\
18.1 & 211.4 & 6.5 & 1.2 & 5.7533 & 4.7555 & 8.42 \\
20.2 & 211.1 & 6.5 & 1.1 & 5.7520 & 4.7525 & 8.43 \\
\bottomrule
\end{tabular}
\end{table}

\begin{figure}
\centering
\includegraphics[width=0.7\linewidth]{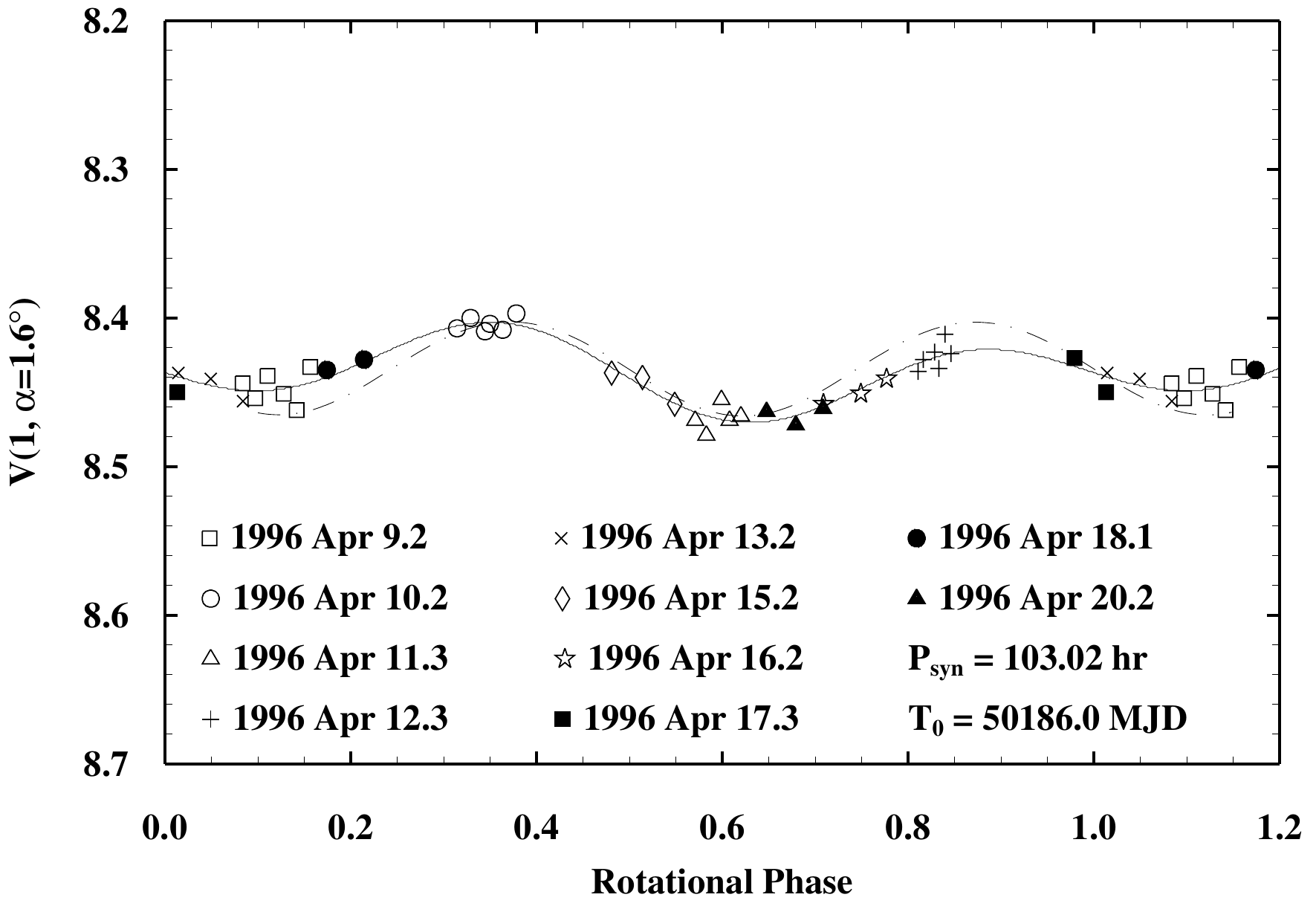}
\caption{Composite visible lightcurve from our April 1996 observations.  Data points at rotational phases 0.0--0.2 are repeated at 1.0--1.2 for clarity. Synthetic lightcurves are superimposed, from a 2nd order Fourier fit to the data (solid line) and from a 1.07:1.0:1.0 ellipsoid model (dash-dotted line).  See text for details.}
\label{fig:lc}
\end{figure}

We observed (617) Patroclus using  the \unit{61}{\cm} Bochum Telescope at the European Southern Observatory in La Silla during 10 nights in April 1996; see \tableref{table:visible} for observational details.  The observations were performed in the Johnson $V$ band with the DLR MKII camera, using a Tektronix 1k x 1k CCD. 

Differential photometry was performed against field stars, which were subsequently absolutely calibrated against observations of standard stars present in near-by fields. Fluxes were extracted through synthetic aperture photometry, using the AstPhot program \citep{Mottola1995}.
The typical $1\sigma$ uncertainties in the differential photometry were well below \unit{0.01}{\text{mag}}; the  typical absolute photometric errors were of the order of \unit{0.02}{\text{mag}} RMS. 

The synodic rotation period was  determined by using the Fourier analysis technique described by \citet{HarrisLupishko1989}.
This technique seeks the period that, in the least-squares sense, best fits the observations.
 In order to remove the effects of changing brightness on subsequent nights due to solar phase effects, or to uncertainties in the nightly absolute photometric calibrations, the method enables grouping of observations (with one group usually comprising observations from one night) and estimates magnitude shifts for each group that minimize the residuals. In our case it became soon apparent that the time spanned by each night of observation was much shorter than the rotation period of Patroclus. As a consequence, when folded into a composite, single-night data stretches would only sporadically overlap with each other, thereby providing little constraint for the solution of the single-night magnitude shifts. For this reason, whenever possible, we grouped the observations in sets of two consecutive nights, which resulted in a total of 6 groups with a good mutual overlap. Subsequently, the magnitudes within each data set were reduced to a reference phase angle (chosen as the phase angle of the observations closest to the average phase angle in each data subset) by using the HG model \citep{HG}
and a $G$ value of 0.12. Because the phase angle change rate at the time of the observations was only of the order of \unit{0.1}{\degree} per day, even a very large deviation of the assumed $G$ value from the actual one would result in a negligible error. 

The data are well described with a Fourier polynomial of 2nd order. 
The resulting best-fit synodic period is  $P_{syn}=103.02 \pm \unit{0.40}{\hr}$.
$P_{syn}$ agrees with the \citet{PatroNew} result of $102.94\pm\unit{0.11}{\hr}$ for the period of the mutual orbit, i.e.\ the rotational variability and the mutual orbit are  synchronized.
See \figref{fig:lc} for a composite lightcurve in which the single-night observations are folded with the lightcurve period, along with the Fourier fit. 
The lightcurve amplitude is found to be $0.070 \pm \unit{0.005}{\text{mag}}$ peak-to-peak.

To  our knowledge, no complete lightcurve of Patroclus has  been published previously, although it was known to have a long period and low amplitude \citep{Angeli1999}.
No signs of multiple periodicity (which might be expected in a binary system) are discernible in our data.
This implies that no mutual events occurred at the time of our observations \citep[consistent with the][model]{PatroNew} and indicates a 
fully synchronized binary system.
The small lightcurve amplitude implies that both components are nearly spherical in shape.
While the data do not allow an unambiguous shape model to be determined,
the observed lightcurve can be reproduced reasonably well assuming two slightly prolate spheroids with axial ratios 1.07:1:1, a realistic photometric function, and the \citeauthor{PatroNew}\ spin axis (see \figref{fig:lc}).

\section{Spitzer observations}
\label{sect:obs}

Patroclus was observed using the InfraRed Spectrograph \citep[IRS;][]{IRS} on board the Spitzer Space Telescope \citep{SST}.
IRS was used in low-resolution spectroscopy mode using the 
modules SL1, LL1, and LL2 
to obtain flux-calibrated spectra in the nominal wavelength range  7.4--\unit{38}{\micron} 
at a relative spectral resolution $\lambda/\Delta\lambda$ between 64 and 128.
The observed flux is practically purely thermal; the reflected component is negligible at these wavelengths \citep{Emery2006}.
The angular separation of Patroclus and Menoetius does not exceed \unit{0.2}{\arcsecond} while the IRS pixel scale is \unit{1.8}{\arcsecond} or coarser; the system was  not spatially resolved.

Time-resolved observations were obtained 
during two consecutive mutual events in June 2006,  referred to as events 1 and 2 in the following. 
In event 1, Patroclus shadowed Menoetius, and vice versa in event 2.
The diameter ratio is only $\sim1.1$ and the mutual orbit is  circular, hence the two events 
produced very similar observable effects.
Both events lasted about \unit{4}{\hr} and
were pure shadowing events (the line of sight toward Spitzer was never obstructed).
Thermal emission from the shadowed surface areas therefore contributes to the observable flux.

\begin{table}
  \caption{Observing geometry at the epoch of our observations.
The heliocentric and Spitzer-centric coordinates are longitude and latitude in J2000 ecliptic coordinates.
All values are constant during our observations to $\pm1$ in the last quoted digit or better.
The absolute visible magnitude equals $H=8.19$ \citep[quoted after][]{SIMPS}; the slope parameter of the phase curve is assumed to be $G=0.15$.
}
\label{table:obsgeometry}
\centering
\begin{tabular}{r|ll}
\toprule
Event:& 1 &2 \\
\midrule
heliocentric distance $r$        & \unit{5.947}{\AU}  & \unit{5.947}{\AU} \\
Spitzer-centric distance  $\Delta $   & \unit{5.95}{\AU}  & \unit{5.98}{\AU}\\
solar phase angle  $\alpha$& \unit{9.80}{\degree}  & \unit{9.77}{\degree}\\
heliocentric coords & \unit{170.8}{\degree}, \unit{+18.03}{\degree} & \unit{170.9}{\degree}, \unit{+18.00}{\degree}\\
Spitzer-centric coords & \unit{160.5}{\degree}, \unit{+18.2}{\degree}& \unit{160.7}{\degree}, \unit{+18.1}{\degree}\\
\bottomrule
\end{tabular}  
\end{table}

\begin{table}
\caption{Start times of our 18 Spitzer observations. 
See \figref{fig:geometry} for the respective system geometry.
}
\label{table:timing}
\centering
\begin{tabular}{rrl|rrl}
  \toprule
    &  Day & Time & & Day & Time \\
    & (June 2006) & (UT) & &(June 2006) & (UT) \\
\midrule
1.0 & 24 & 18:40 & 2.0&  26& 10:42\\
1.1 & 24 & 21:54 & 2.1&  26& 23:22\\
1.2 & 24 & 22:47 & 2.2&  27& 00:24\\
1.3 & 24 & 23:54 & 2.3&  27& 01:31\\
1.4 & 25 & 00:41 & 2.4&  27& 02:19\\
1.5 & 25 & 01:47 & 2.5&  27& 03:29\\
1.6 & 25 & 02:49 & 2.6&  27& 04:24\\
1.7 & 25 & 04:12 & 2.7&  27& 05:55\\
1.8 & 25 & 05:24 & 2.8&  27& 06:52\\
\bottomrule
\end{tabular}
\end{table}

\begin{figure}
  \centering
\includegraphics[width=\linewidth]{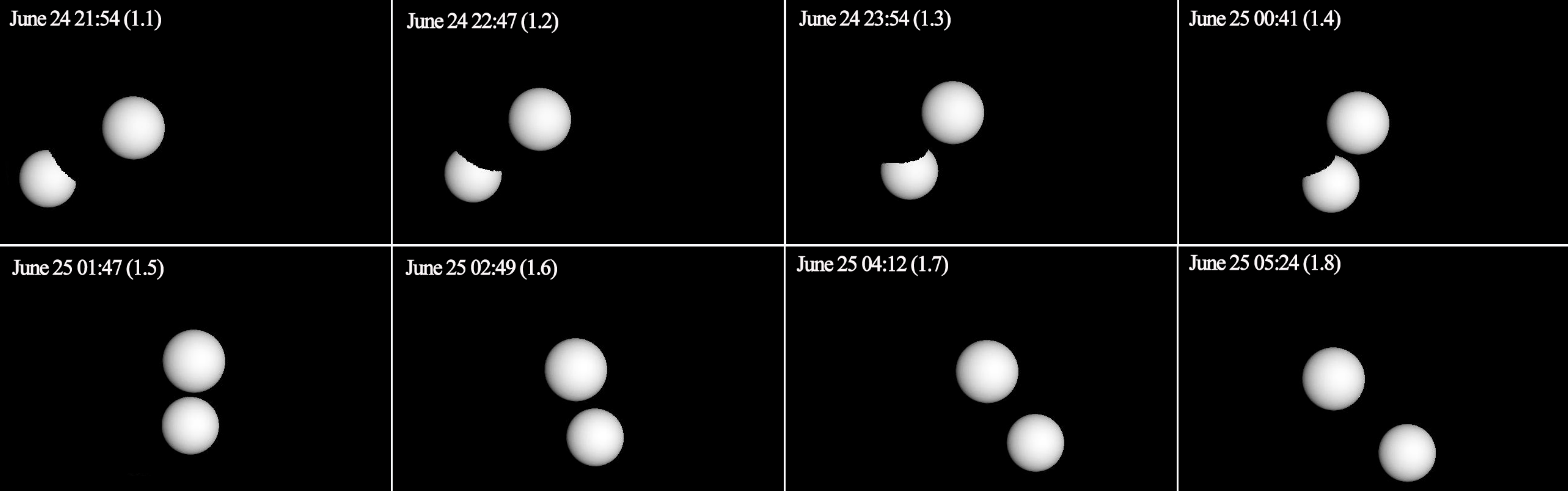}

\medskip

\includegraphics[width=\linewidth]{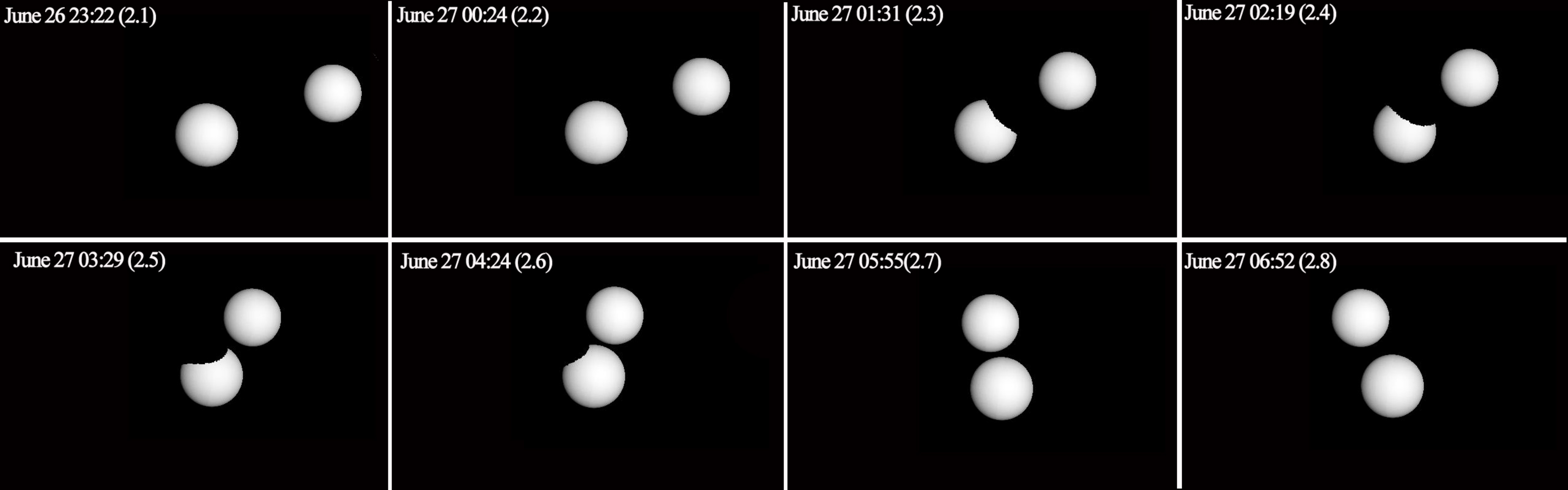}
  \caption{Geometry of the system's mutual orbit during our Spitzer observations as seen from Spitzer, generated using the \citet{PatroNew} orbit model.  
Each frame corresponds to one of our observations 1.1--1.8 (top two rows) and 2.1--2.8 (lower two rows), see \tableref{table:timing}.  Observations 1.0 and 2.0 are not plotted; at this scale, only one component would be displayed.
Each frame is centered on Patroclus; its 
angular diameter  is $\sim\unit{12.4}{\text{milli arcsec}}$. 
During event 1, Menoetius is shadowed by Patroclus; and vice versa during event 2.  The line of sight toward Spitzer was never obstructed.
The event timing was shifted somewhat relative to the \citeauthor{PatroNew}\ prediction (within the uncertainties) to match the best-fit eclipse timing found in \sectref{sect:results}.
The system is eclipsed during observations 1.1--1.4 and 2.3--2.6, respectively.
}
  \label{fig:geometry}
\end{figure}

A total of 18 thermal-infrared spectra of Patroclus were obtained, nine per event, referred to as 1.0--1.8 and 2.0--2.8, respectively.  See \tableref{table:obsgeometry} for the observing geometry during the events. 
Observation start times are given in \tableref{table:timing}; 
each observation lasted \unit{$\sim6$}{\minute}.
As can be seen in \figref{fig:geometry}, our observations 
provide good temporal coverage of both events and their aftermath.
The \citet{Marchis2006} orbit model was used to predict the time of the events several months ahead.
Observations 1.0 and 2.0 were performed well before the start of the events, in order to establish a flux baseline.

To prevent
flux from 'spilling over' the edge of the slit, 
IRS spectroscopy targets must be accurately centered into the slit.
The projected width of the LL slit is above \unit{10}{\arcsec}, while that of the SL1 slit is \unit{3.7}{\arcsec}, comparable to the width of the point-spread function (PSF) in the respective wavelength range.  For comparison: the pixel scale is \unit{1.8}{\micron} in the SL modules and \unit{5.1}{\micron} in the LL modules.
 ``Blind'' telescope pointings have a $1\sigma$-accuracy of \unit{0.5}{\arcsec} and are therefore adequate to center sources of well known position (such as Patroclus) in the wide LL slits but risk placing a significant fraction of the PSF outside the SL1 slit.
Instead, small ``spectral maps'' were created for the SL1 observations, with three small steps perpendicular to the slit, each offset by \unit{2}{\arcsec} (roughly half the slit width). 
This allowed us to estimate the target offset from the slit center and to correct for the effects thereof in the data analysis (see below).

The usual ``nod'' strategy was used, i.e.\ the target was placed  at about 33 and \unit{66}{\%} of the slit length in each module in order to enable subtractive correction for diffuse background emission (``sky background'').

The data obtained were reduced  using the methods described in \citet[sect.\ 3]{Emery2006}.
Briefly, the Spitzer Science Center (SSC) receives the data from the telescope and runs it through a reduction pipeline that converts raw data into 2-D spectral flux images (electrons/s) while flagging bad pixels, correcting known stray light artifacts, and performing flat field corrections.  The output is termed Basic Calibrated Data (BCD), which is described in detail in the IRS Data Handbook (\url{http://ssc.spitzer.caltech.edu/irs/dh/}).  The data presented in this paper result from the BCD pipeline version S16.1.0.

The BCD files contain background emission, mostly from the zodiacal cloud, which is removed by subtracting the two nod positions from one another for each observation.  Extraction of the data to 1-D spectra essentially follows that of ground-based spectral data, where the object is identified in each row and the dispersion mapped across the chip.  The signal ($e^-/\second$ at this point) is the sum of the pixels within the chosen extraction width at each wavelength.  Fractional pixels are included assuming the flux is evenly distributed across the individual pixel.  The SSC provides a wavelength map of the chip for each spectral order.  Since Spitzer is diffraction limited, the PSF, and therefore optimal extraction width, varies with wavelength.  We use the same extraction width as recommended by the SSC: $\sim 2$ x FWHM (\unit{14.4}{\arcsec}, \unit{21.68}{\arcsec}, and \unit{36.58}{\arcsec}  at \unit{12}{\micron}, \unit{16}{\micron}, and \unit{27}{\micron}, respectively) and varying linearly with wavelength.  Because of the varying PSF width and several other artifacts, the extracted spectral shape for each order is known to be slightly incorrect.  The SSC corrects (''tunes'') the final spectral shape using a 5th order polynomial fit to the ratio of stellar calibrator observations to models of their spectral shape \citep{Decin2004}.  These ratios are also used for final flux calibration (i.e., converting from $e^-/\second$ to Jy).  We perform basically the same procedure, with the exception that we use different calibrator stars, and we do not fit a polynomial, but rather apply the average ratio of the stars to the model directly.  For the data presented here, we used del UMi, HR 2194, and HR  7891 for the SL modules and HD173511, del UMi, and HR 7341 for the LL modules.  The different orders (SL1, LL2, and LL1) overlap slightly, and we scale the different orders relative to each other using these overlap regions.  We then rescale the entire spectrum to the mean of the individual scale factors.  The different low resolution orders generally agree within $<\unit{7}{\%}$.  

\begin{figure}
  \centering
\includegraphics[width=0.6\linewidth]{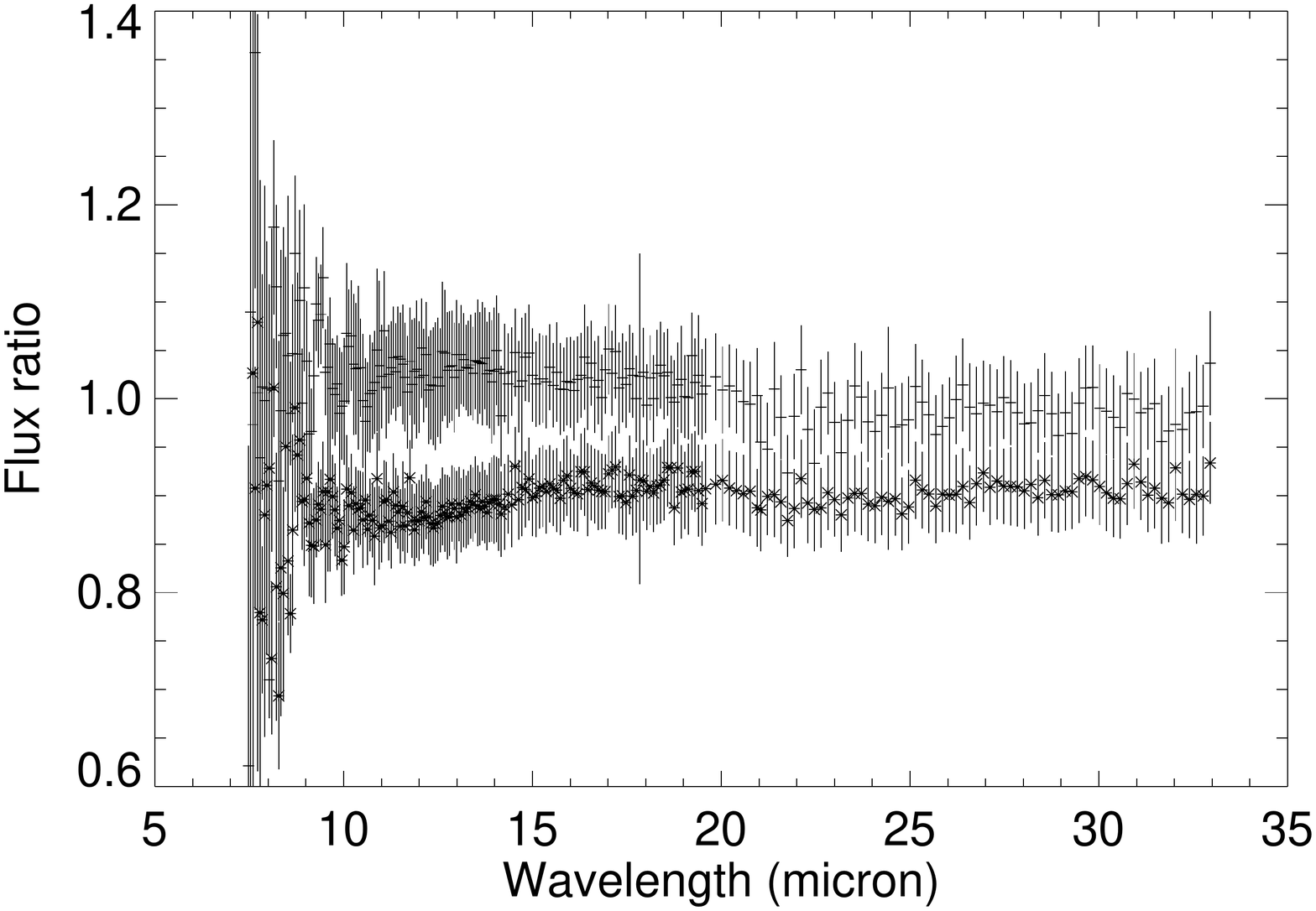}
\medskip
\includegraphics[width=0.6\linewidth]{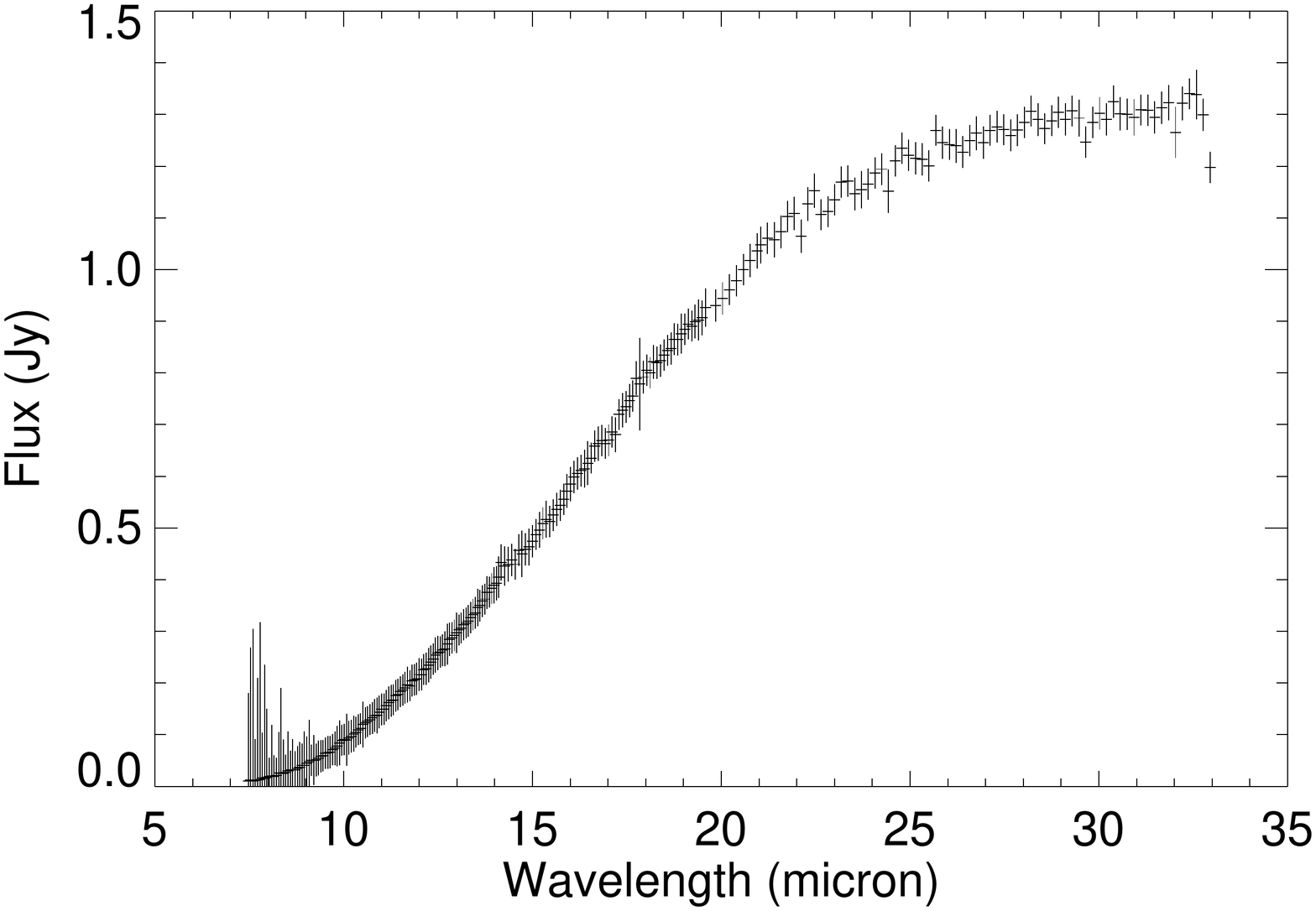}
  \caption{Top: Ratio of two measured thermal spectra relative to pre-event measurement 1.0. Observation 1.2 (bottom) was taken during the eclipse, observation 1.7 (top) in the aftermath. Note the difference in flux level and spectral slope, indicative of the eclipse-induced temperature drop.
Bottom: Spectrum 1.0 in absolute flux units for comparison.}
  \label{fig:fluxratios}
\end{figure}

Observations at some wavelengths are compromised by defective detector pixels and are disregarded in the following, 
e.g.\ all wavelengths above \unit{33}{\micron}.
Systematic flux uncertainties due to, e.g., the residual effect of source mis-centering or uncertainties in the absolute flux calibration 
are estimated to be \unit{3}{\%} and added
to the statistical uncertainty.
The total flux uncertainty is dominated by the systematic uncertainty except for the low fluxes on the Wien slope at the shortest wavelengths.
Final fluxes are provided in the Online Supplementary Material.
See \figref{fig:fluxratios} for a  plot of representative flux ratios during event 1.

\section{Emissivity features}
\label{sect:features}

The observed spectra contain slight spectral features due to silicates within the wavelength ranges  10--\unit{12}{\micron} and 18--\unit{22}{\micron}.
Emissivity spectra are calculated by dividing the measured flux spectrum by the modeled thermal continuum.
Since we are not interested in smooth spectral slopes but  in discrete features, the specific choice of model continuum 
is uncritical at this point.
We see no reliable spectral variation among the 18 observations and therefore show the grand average.  

\begin{figure}
  \centering
\includegraphics[angle=0,width=0.6\linewidth]{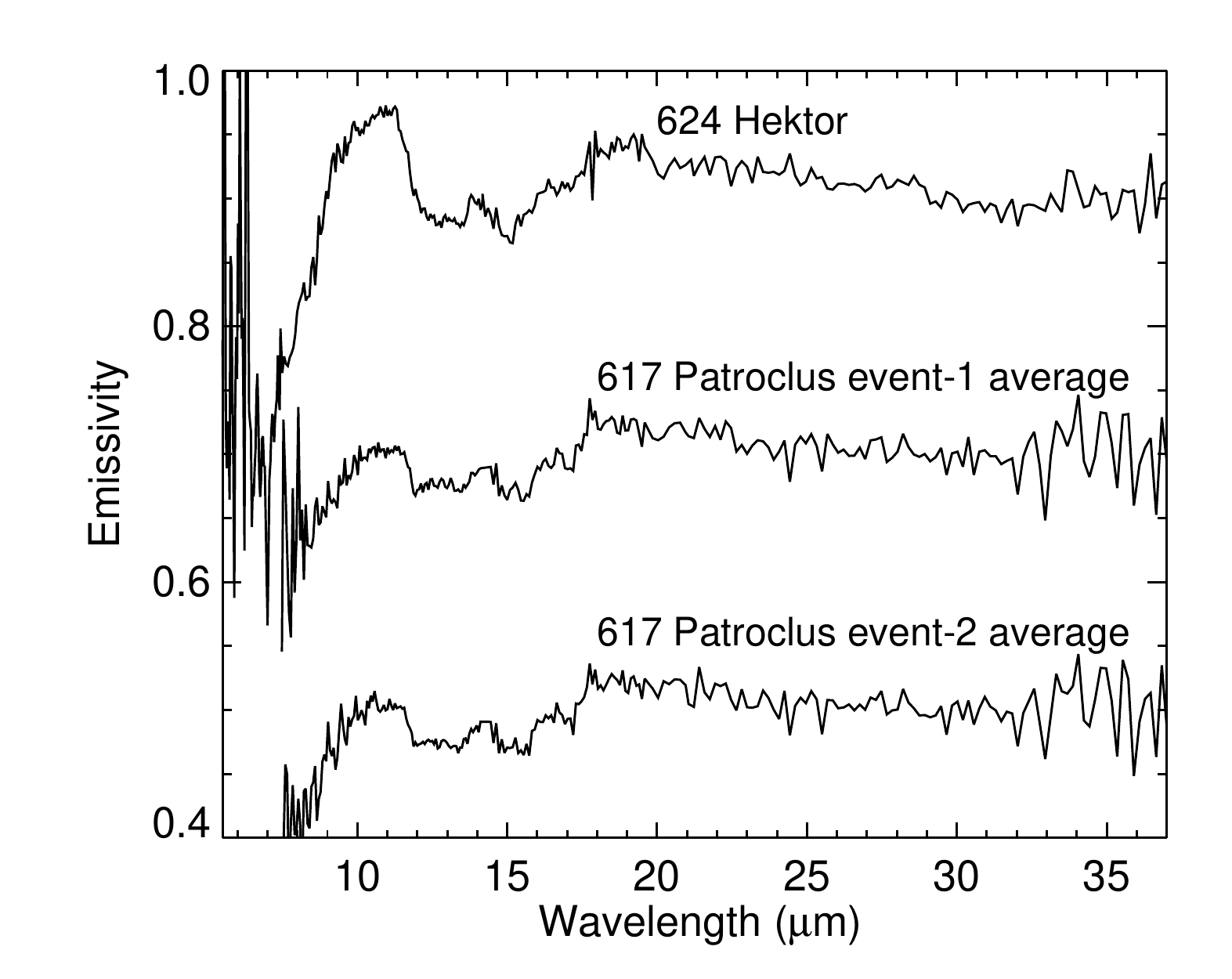}
\caption{Emissivity spectra of Patroclus (see text) and Hektor \citep{Emery2006}. Spectra are shifted vertically for clarity.}
  \label{fig:emissivity}
\end{figure}

Our final emissivity spectrum result for the Patroclus system is shown in \figref{fig:emissivity} and compared to that of the Trojan asteroid (624) Hektor \citep{Emery2006}.
Features near 10 and \unit{20}{\micron} are clearly visible; small scale (single channel) structure in the Patroclus spectra, e.g.\ near \unit{25}{\micron}, are residual calibration artifacts.

\citeauthor{Emery2006}\ conclude that the features near 10 and \unit{20}{\micron} in Hektor (and two other Trojan asteroids) are due to the presence of fine-grained ($<$ few \micron) silicates, likely in either a very under-dense (''fairy-castle'') structure or embedded in a matrix material that is fairly transparent at these wavelengths (e.g., certain organic materials).  Since the broad emissivity peaks in the Patroclus spectrum occur at the same wavelengths and with very similar shapes to those in the spectrum of Hektor, it is reasonable to infer similar mineralogy for the surface of Patroclus.  The features, particularly the one near \unit{10}{\micron}, are more subdued in the spectrum of Patroclus than that of Hektor \citep[or the other two Trojans presented in][]{Emery2006}.  This is most likely due to differences in grain size or packing state of the surface, two factors which strongly affect the shape of spectra at these wavelengths, but could also imply different mixing ratios of silicates to opaque material.  

In order to avoid biases in thermal modeling, the wavelengths at which these features occur 
(10--\unit{12}{\micron} and 18--\unit{22}{\micron})
were disregarded in the thermal analysis, leaving 178 data points per observation.

\section{Thermophysical modeling}
\label{sect:TPM}

Analyzing thermal-infrared observations of asteroids requires a thermal model to calculate expected fluxes.
To this end, the temperature distribution is calculated based on the observing geometry and assumed physical properties, then the observable model flux is calculated by integrating the Planck function over the visible surface.

Thermal fluxes depend on a number of physical properties, among them the effective diameter $D$, geometric albedo \pv, shape, thermal inertia $\Gamma$, spin period $P$,  spin-axis orientation, and amount of surface roughness.
Typically, the absolute visible magnitude $H$ 
is known, such that \pv\ and $D$ are linked through  \citep[see][Appendix A, and references therein]{PravecHarris2007}
\begin{equation}
\label{eq:dpvh}
D=10^{-H/5} \frac{\unit{1329}{\km}}{\sqrt{\pv}},
\end{equation}
reducing the number of free parameters by one.

The surface temperature distribution is governed by $\Gamma$ (together with the spin rate):
An object with zero thermal inertia (or vanishing spin rate) would display a prominent temperature difference between the hot sub-solar point and the $T=0$ night side, whereas increasing $\Gamma$ (or spin rate) shifts the temperature maximum toward the afternoon side and, for observations at low phase angles, reduces the apparent color temperature, since solar energy is stored and re-emitted at night.

Surface roughness, e.g.\ due to cratering,  causes thermal-infrared beaming: 
relative to a smooth surface, a rough surface displays 
more surface elements facing the Sun, even close to the limb, leading to a preferentially sunward emission and, for observations at low phase angle, a higher color temperature.
Also, surface elements inside an indentation can radiatively exchange energy, leading to mutual heating. 

Simple thermal models in which highly idealized assumptions are made on shape, thermal inertia, and surface roughness have enjoyed widespread usage over the past decades and have been shown to allow robust estimates of diameter and albedo in many cases \citep[see][for a recent overview]{Harris2006}.

Deriving the thermal inertia, on the other hand, requires more realistic thermophysical modeling, and in particular some knowledge of the object's spin state, which is currently available for a small (but growing) number of asteroids only.
Also, given the notorious difficulty of ground-based mid-IR observations,
complex models with a large number of free parameters are not frequently warranted by the data quality.
Nevertheless, detailed thermophysical models (TPMs) 
 in which shape, spin state, thermal inertia, and surface roughness are explicitly taken into account,
have been developed
\citep[e.g.][]{Spencer1990,LagerrosI,LagerrosIV,Mueller2007,Delbo2009}
and have been used successfully to analyze mid-IR observations of asteroids
\citep[e.g.][]{MuellerLagerros1998,Harris2005,Harris2007,Mueller2007}.

Since the thermal skin depth, a measure of the penetration depth of the  heat wave, is typically in the \centi\metre-range, lateral heat conduction can be neglected.
We also assume, as is common practice, that all relevant parameters are constant with depth and temperature.%
\footnote{
\label{footnote:depth}%
In the case of lunar regolith, thermal parameters are actually empirically known to depend on both temperature and depth
\citep[see, e.g.,][and references therein]{Jones1975}.
It is, however, unclear how to apply lunar results to asteroid regolith, which may be substantially different.
Thermal-inertia results derived from TPMs should be thought of as effective averages over the relevant length scales.}
It is then practical to use dimensionless units 
\citep[see][]{Spencer1989}
for time $t^\prime$, depth $X$, and temperature $T^\prime$.
The only free parameter in the heat conduction problem is then the thermal parameter $\Theta$, as defined by \citeauthor[][]{Spencer1989},
which is proportional to $\Gamma$.
The equations governing thermal conduction into the subsoil are
\begin{align}
\label{eq:diffusion}
\frac{\partial}{\partial t^\prime}T^\prime\left(X,t^\prime\right) &= \frac{\partial^2}{\partial X^2}T^\prime\left(X,t^\prime\right) \\
\label{eq:bc:surface}
T^\prime(0,t^\prime)^4 &= \mu_S(t^\prime) + \Theta\frac{\partial}{\partial X}T^\prime(0,t^\prime) \\
\label{eq:bc:infty}
\lim_{X\to\infty} \frac{\partial}{\partial X}T^\prime(X,t^\prime) &= 0
\end{align}
where $\mu_S$ is the cosine of the local solar zenith distance (clipped to be $\geq 0$).
\Eqref{eq:diffusion} describes heat diffusion within the subsoil, \eqref{eq:bc:surface} the surface boundary condition (thermally emitted energy equals absorbed solar energy plus the net heat flux from the subsoil), \eqref{eq:bc:infty} ensures the regularity of the solution.
See, e.g., \citet[][Sect.\ 3.2.2]{Mueller2007} for a more detailed discussion.

Surface roughness can be modeled \citep[following][]{Spencer1989,Spencer1990} by multiplying surface temperatures as calculated from Eqns.\ \ref{eq:diffusion}--\ref{eq:bc:infty} with a factor $\eta^{\prime-1/4}$ ($\eta^\prime$ is called $\beta$ in the quoted papers).   $\eta^\prime < 1$ leads to elevated apparent color temperature.
Increasing roughness leads to decreasing values of $\eta^\prime$, a smooth surface would display $\eta^\prime=1$. \citet{Spencer1989} derive $\eta^\prime\sim0.72$ for the lunar surface.

\subsection{Binary TPM (BTPM)}
\label{sect:BTPM}

We describe a new Binary TPM (BTPM). It is based on the  \citet{Spencer1990} TPM, but includes
the effect of mutual events (eclipses and occultations) on thermal fluxes.  
Our BTPM is limited to fully synchronized binary systems with a circular mutual orbit, such as Patroclus-Menoetius.
Such objects are at rest in a co-rotating coordinate system.
We can therefore use a single-body TPM to model the binary system, using a shape model which consists of two disjoint parts.

In contrast to most TPMs, which assume a globally convex shape, 
surface elements can now shadow one another and obstruct the line of sight toward the observer.
In principle, some facets can also radiatively exchange energy, similar to mutual heating inside craters. 
That latter effect is not included in the model. While it is important in the thermal modeling of globally non-convex bodies, it is negligible for eclipse observations at low solar phase angle.

When generating the binary shape model, 
both components are modeled as a mesh of triangular facets.  Arbitrary component shapes are allowed by the model code; spheres and ellipsoids are used for this study.
Each facet is checked for potential shadowers, i.e.\ other facets which appear above the local horizon. The list of potential shadowers along with details on the mutual viewing geometry is stored as part of the shape model \citep[see][Sect.\ A.1.1 for details]{Mueller2007}, ensuring that these calculations do not need to be repeated for each BTPM run.

Since mutual heating is neglected, 
the observable emission 
can be calculated facet-by-facet. Mutual heating would introduce  an interdependence of surface temperatures and add to the complexity.  
Temperatures are calculated by numerically solving  Eqns.\ \ref{eq:diffusion}--\ref{eq:bc:infty}. 
Eclipses are taken into account by setting $\mu_S$, 
the insolation term in \eqref{eq:bc:surface},
to zero when the facet in question is eclipsed by another facet.
Facets are considered eclipsed when their midpoint is shadowed (no partial shadowing of individual facets).
Care must be taken to choose a sufficient time resolution, so as to provide appropriate resolution of eclipse events, which are short compared to the rotation period.

\section{NEATM analysis}
\label{sect:NEATM}

As a first step, the obtained spectra (minus the location of emissivity features) were analyzed using the NEATM \citep{Harris1998}.  For the purposes of this study, it is useful to think of the NEATM as a simplified TPM in which the shape is assumed to be spherical and thermal inertia is not directly modeled (therefore the spin state is irrelevant). 
It contains a model parameter $\eta$, which modifies model surface temperatures in the  same way as our BTPM's $\eta^\prime$ (temperature $\propto\eta^{-1/4}$),
but is used for an effective description of both surface roughness \emph{and} thermal inertia.
For observations at low phase angles, increasing thermal inertia lowers surface temperatures  and therefore increases $\eta$ (all other parameters kept constant).

The NEATM has been used widely to analyze thermal-IR observations of asteroids and is known to provide diameter results which are in generally good agreement with diameter determinations obtained using other methods.  Additionally, conclusions on thermal properties such as thermal inertia can be drawn from the best-fit $\eta$ value.
See, e.g., \citet{Harris2006} for a recent overview.

\begin{table}
\caption{NEATM fit to our Spitzer data: best-fit $\eta$, diameter $D$, and geometric albedo \pv. Uncertainties were derived through a straight-forward Monte-Carlo analysis (mean and average of fits to 300 random spectra per observations, normally distributed about the measured data) and account for the statistical uncertainty, only.}
\label{table:NEATM}
\begin{tabular}{rlll|rlll}
Obs.\ & $\eta$ & $D$(\km) & \pv (\%)&
Obs.\ & $\eta$ & $D$(\km) & \pv (\%)\\
\hline
1.0 & $0.845\pm0.004$ & $146.8\pm  0.5$ & $4.34\pm0.03$ &
2.0 & $0.839\pm0.004$ & $148.5\pm  0.5$ & $4.24\pm0.03$ \\
1.1 & $0.852\pm0.004$ & $143.5\pm  0.6$ & $4.54\pm0.04$ &
2.1 & $0.856\pm0.004$ & $144.9\pm  0.6$ & $4.45\pm0.04$ \\
1.2 & $0.864\pm0.004$ & $141.3\pm  0.5$ & $4.68\pm0.04$ &
2.2 & $0.855\pm0.005$ & $143.4\pm  0.6$ & $4.55\pm0.04$ \\
1.3 & $0.860\pm0.006$ & $140.9\pm  0.7$ & $4.71\pm0.05$ &
2.3 & $0.878\pm0.005$ & $141.5\pm  0.5$ & $4.67\pm0.04$ \\
1.4 & $0.856\pm0.004$ & $143.2\pm  0.5$ & $4.56\pm0.04$ &
2.4 & $0.895\pm0.006$ & $138.9\pm  0.6$ & $4.85\pm0.04$ \\
1.5 & $0.848\pm0.004$ & $146.7\pm  0.6$ & $4.34\pm0.04$ &
2.5 & $0.863\pm0.004$ & $133.6\pm  0.4$ & $5.24\pm0.04$ \\
1.6 & $0.827\pm0.004$ & $144.1\pm  0.6$ & $4.50\pm0.04$ &
2.6 & $0.891\pm0.004$ & $142.5\pm  0.5$ & $4.60\pm0.04$ \\
1.7 & $0.835\pm0.004$ & $146.1\pm  0.5$ & $4.38\pm0.03$ &
2.7 & $0.875\pm0.004$ & $144.1\pm  0.6$ & $4.50\pm0.04$ \\
1.8 & $0.851\pm0.004$ & $147.2\pm  0.5$ & $4.32\pm0.03$ &
2.8 & $0.861\pm0.004$ & $143.3\pm  0.6$ & $4.55\pm0.04$ \\
\end{tabular}
\end{table}

We fitted the NEATM to each of our 18 observations, see \tableref{table:NEATM} for the results.
Since the NEATM does not include the effects of eclipses, it is not applicable during and just after eclipses.
By averaging over observations 1.0, 1.7, 1.8, and 2.0--2.2 (see \figref{fig:geometry} and \tableref{table:timing}), 
we obtain $D\sim\unit{146}{\km}$ and 
$\eta\sim0.85$.  The $\eta$ value indicates a relatively low thermal inertia and significant surface roughness, but does not provide a quantitative constraint on either.

\begin{figure}
  \centering
\includegraphics[width=0.6\linewidth]{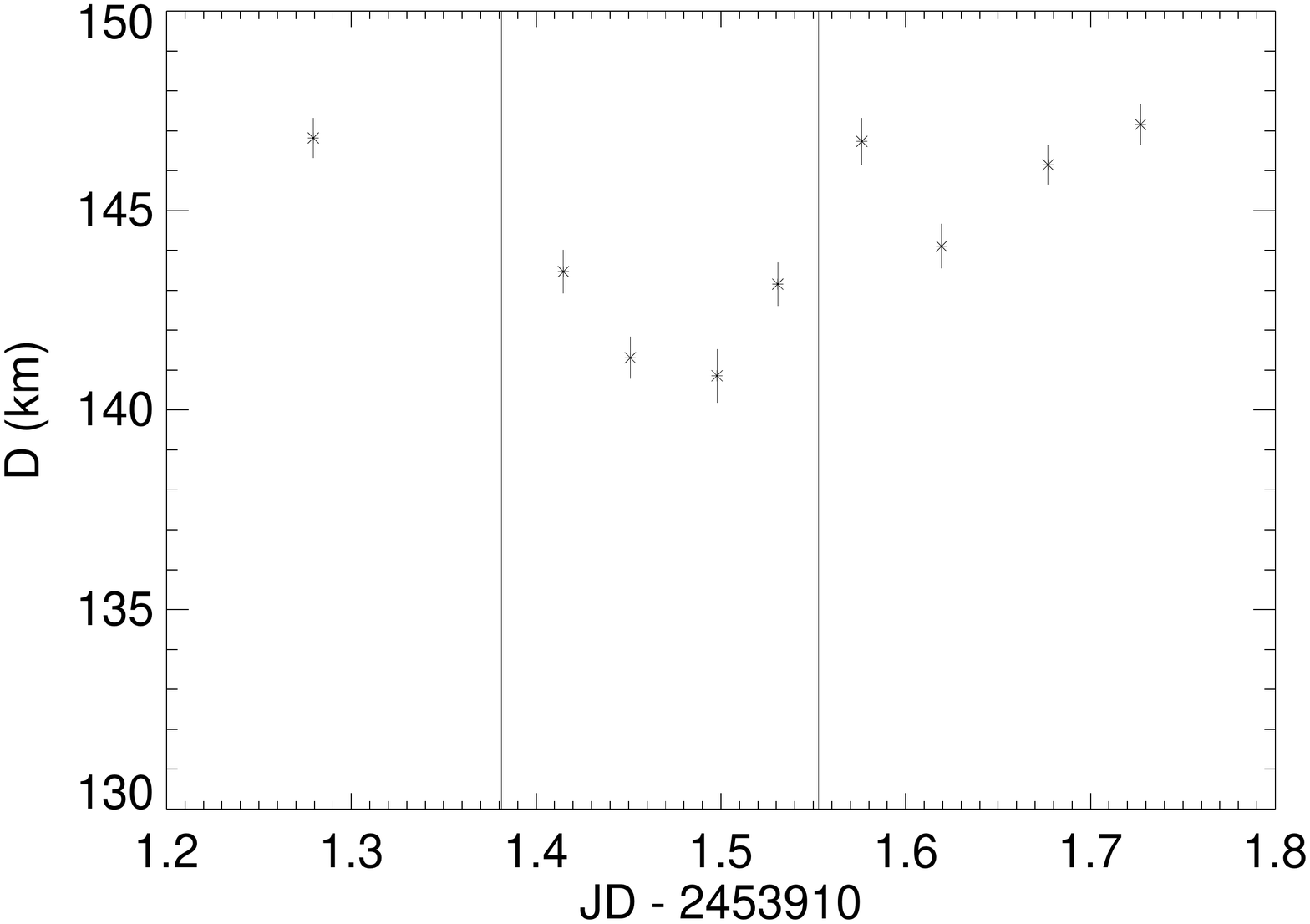}

\includegraphics[width=0.6\linewidth]{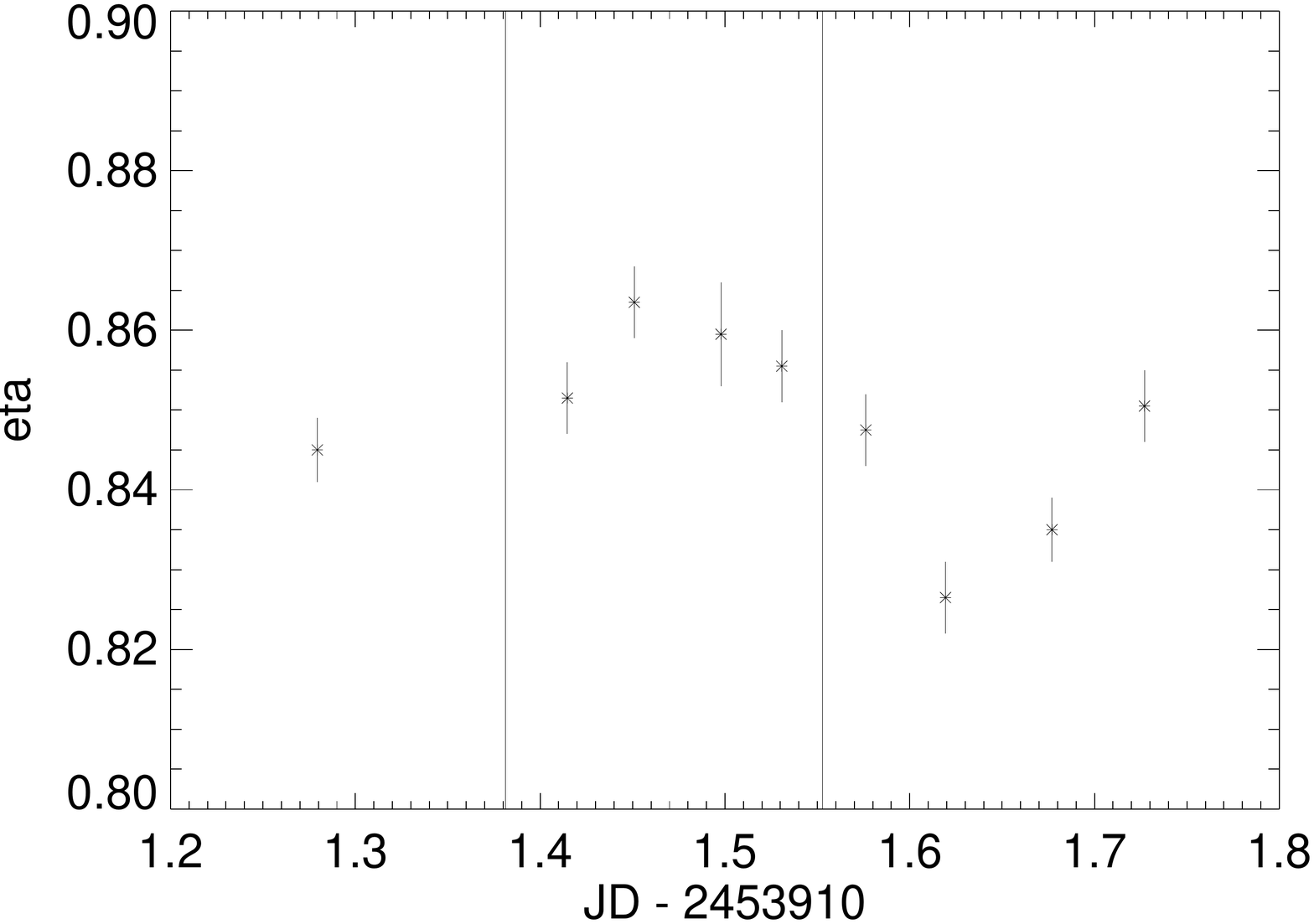}
  \caption{NEATM fits to observations 1.0--1.8 (see \tableref{table:NEATM}). Above: Best-fit effective diameter against observing time. Below: Best-fit $\eta$ against observing time. The vertical lines denote beginning and end of the eclipse event. The error bars are purely statistical.}
  \label{fig:NEATM}
\label{lastfig}
\end{figure}

While the NEATM takes no direct account of shadowing, the reduced total flux and the reduced  temperature during and just after the eclipse events should be reflected in the NEATM results.
Indeed, as can be seen in \figref{fig:NEATM}, 
there is a  clearly recognizable dip in NEATM diameter during event 1 (results for event 2 are qualitatively identical), which reflects the event-induced flux drop.
Simultaneously, the best-fit $\eta$ rises, corresponding to a lower apparent color temperature
due to the eclipse-induced cooling of the shadowed parts.

While no quantitative conclusion on thermal inertia can be drawn from the NEATM analysis, our results indicate that we have indeed observed the thermal response to eclipse events.

\section{BTPM analysis}

\subsection{Fitting technique}
\label{sect:modeling}

In this section, we describe fits to our Spitzer data (denoted as $d_i$ for the data and $\sigma_i$ for their uncertainties)
using the new BTPM described in \sectref{sect:BTPM}. 
Variable parameters are  $\Gamma$, $\eta^\prime$, area-equivalent system diameter $D_\Area$, and the time of the events
(eclipse time predictions from the orbit model are currently uncertain by $\pm$~a few hours).
To this end, synthetic lightcurves were generated using the observing geometry given in \tableref{table:obsgeometry} (separately for each event) and
 for a large range of $\Gamma$ and $\eta^\prime$ values at those wavelengths $\lambda_i$ at which Spitzer data have been obtained. The positions of emissivity features (10--12  and 18--\unit{22}{\micron}, see \sectref{sect:features}) were disregarded.
Model fluxes $m_i(\lambda_i)$  were calculated for 1000 time points per revolution, corresponding to a time resolution of $\sim\unit{6}{\minute}$ or roughly the length of one Spitzer observation. This provides more than 40 time points across the duration of each $\sim\unit{4}{\hour}$ event.
The rough rotational phase of the events was determined from visual inspection of the synthetic lightcurves, the exact timing was determined by fitting a time offset $\Delta t$.
All model calculations were performed assuming $H=8.19$, 
$\pv=0.0433$, and $D_\Area=D_0 =  \unit{147}{\km}$.
Fluxes are assumed to be proportional to $D_\Area{}^2$: $m_i(\lambda_i, D_\Area) = x\ m_i(\lambda_i, D_0)$ with $x=(D_\Area/D_0)^2$.%
\footnote{Implicitly, we neglect the albedo dependence of thermal fluxes: varying $D_\Area$ with constant $H$ varies \pv. Temperatures depend on the absorptivity, which is a strong function of \pv\ for high albedo, but a weak function for low \pv.  
Since Patroclus' albedo is very low, we reckon this approximation is uncritical \citep[see also][]{HarrisHarris}.}
For each considered set of $\Gamma$, $\eta^\prime$, and $\Delta t$ the best-fit $x$ is found by minimizing (through linear regression)
\begin{equation}
\label{eq:fitting:scale}
\chi^2 = \sum_i \frac{\left[x\ m_i(\lambda_i) - d_i(\lambda_i) \right]^2}{\sigma_i(\lambda_i)^2}.
\end{equation}
The corresponding best-fit diameter equals $D_\Area = \sqrt{x}\ \unit{147}{\km}$.

After initial runs over wider and coarser grids, 
an equidistant grid was used with a step width of \unit{0.5}{\TIunit} in $\Gamma$, 0.0025 in $\eta^\prime$, and \unit{1}{\minute} in $\Delta t$.
Data were analyzed separately for each event.
For each combination of $\Gamma$, $\eta^\prime$, and $\Delta t$, 
the best-fit $x$ and corresponding $\chi^2$ were determined 
 and stored.
The grid was then searched for  that combination of $\Gamma$, $\eta^\prime$, and $\Delta t$  which led to the global minimum in $\chi^2$. 

In order to study the accuracy of the results, a Monte-Carlo technique was employed. 
For each observation, 5000 random (''noisy'') spectra were generated;
for each wavelength the distribution of the 5000 random points 
is Gaussian, with a mean value equal to the measured flux and a standard deviation equal to the flux uncertainty.
Best-fit BTPM parameters were determined for each  set of nine random spectra; 
mean value and standard deviation of the resulting values were adopted as 
nominal value and statistical uncertainty, respectively.
This error analysis does not yet account for the uncertainty in the axis of the mutual orbit \citep[$\lambda = 241.37\pm\unit{0.33}{\degree}$, $\beta=-57.35\pm\unit{0.36}{\degree}$;][]{PatroNew}.
Eclipse depth depends sensitively on the sub-solar latitude and therefore on the orbital axis, thus the axis uncertainty (while small) could be expected to have a significant impact on the accuracy of our thermal-inertia result.
We  repeated the Monte-Carlo BTPM analysis assuming 
 the axis within the quoted range of uncertainty  
that leads to the largest change in sub-solar latitude:  $\lambda=\unit{241.70}{\degree}$, $\beta=\unit{-57.71}{\degree}$.
The offset axis leads to a somewhat increased sub-solar latitude, implying shallower eclipse events, i.e.\ a smaller area fraction of the eclipsed component is shadowed.

Additional simulations were performed in which the components were assumed to be prolate spheroids with aligned long axes (nominal orbit solution).
Axis ratios (identical for the two components) of up to 1.09 were tried; larger ratios would be inconsistent with the low visible lightcurve amplitude \seesect{sect:visible}.

\subsection{Results}
\label{sect:results}

\begin{table}
\caption[Best-fit BTPM parameters for events 1 and 2.]{Best-fit BTPM parameters for events 1 and 2.
The two lines of results per event refer to the nominal orbit solution and that offset by $1\sigma$ (see text), respectively.
Syzygy times are as seen on Spitzer.
There are $9\times178=1602$ data points per event and  four fit parameters, hence a best-fit \chitwo\ in the low thousands is to be expected.}
\label{table:results}
  \centering
  \begin{tabular}{r|rrrrr}
\toprule 
& Syzygy
& \chitwo 
& $\Gamma$ & $\eta^\prime$ & $D_\Area$ \\
& ($JD-2453910$) & & (\TIunit) & & (\km) \\
\midrule
Event 1 & $1.466\pm 0.003$   & $2543\pm84$ & $20.7\pm3.8$ & $0.762\pm0.014$ & $145.7\pm0.3$ \\
        & $1.470\pm 0.002$   & $2586\pm84$ & $7.6\pm1.7$ & $0.814\pm0.008$ & $146.0\pm0.3$ \\
Event 2 & $3.624\pm0.001$ & $3566\pm102$ & $6.4\pm0.9$ & $0.838\pm0.005$ & $143.4\pm0.3$ \\
       &  $3.624\pm0.001$ & $3847\pm109$ & $5.1\pm0.8$ & $0.845\pm0.004$ & $143.1\pm0.3$ \\
\bottomrule
  \end{tabular}
\end{table}

BTPM results are reported in \tableref{table:results} for both the nominal and the offset axis. Results from the ellipsoidal-shape model runs are indistinguishable and are not reported.

The best-fit diameter values resulting from the four BTPM analyses and the NEATM analysis ($\sim\unit{146}{\km}$; see \sectref{sect:NEATM}) are mutually consistent at the \unit{1}{\%} level and average to \unit{144.6}{\km}.

The time of the syzygy (maximum eclipse depth, when the two components and the Sun are aligned) was fitted to the data and is consistent with the \citet{PatroNew} model prediction within the uncertainty of $\pm$ a few hours.
The times given in \tableref{table:results} are for a Spitzer-centric system; subtract the one-way light-time for the corresponding times at Patroclus.

Model runs with the offset axis lead to lower thermal-inertia results in both cases.  This is expected since the offset reduces the eclipse depth:
The leading-order effect of increasing thermal inertia or decreasing eclipse depth is to reduce the eclipse-induced flux drop; hence underestimated eclipse depth leads to underestimation of thermal inertia.
Final error estimates are obtained by adding in quadrature the statistical uncertainty in the nominal solution to the difference between the nominal and offset-axis solutions; our final results are  $\Gamma=21\pm\unit{14}{\TIunit}$ (event 1) and $\Gamma=6.4\pm\unit{1.6}{\TIunit}$ (event 2).

While our two thermal-inertia results differ by a factor $>3$, they 
are mutually consistent given the uncertainties. 
Note, however, that neither result is representative of the system as a whole, 
but is dominated by the respective shadowed region.
A difference between the two thermal-inertia results
would therefore imply inhomogeneous regolith properties and might hint at component-to-component differences.
While tantalizing, our data do not allow this conclusion to be drawn at a statistically significant level.
We estimate the system-average thermal inertia to be between 5 and \unit{35}{\TIunit} or $20\pm\unit{15}{\TIunit}$, where the scatter may be partially due to surface inhomogeneity.

Our thermal-inertia results depend sensitively on the assumed spin axis, as evidenced by our experiment with the offset axis and, more drastically, by our experience with preliminary orbit models.  
E.g., in a preliminary data analysis based on a preliminary orbit model \citep[][Sect.\ 6.8]{Mueller2007},  a best-fit  $\Gamma$ of $\sim\unit{90}{\TIunit}$ was obtained. The uncertainty introduced by the spin-axis uncertainty was not studied in the previous analysis.
We point out, however, that the new orbit model fits the Spitzer data 
much better: \chitwo\ is now reduced by a factor $>3$ and is now in the expected range given the number of data points.
Also, the nominal spin axis fits the Spitzer data better than that offset by $1\sigma$,
inspiring further trust in the presented results.

The best-fit  $\eta^\prime$  ranges from 0.76 to 0.85, between the lunar value of $\eta^\prime\sim0.72$ \citep{Spencer1989} and the NEATM best-fit $\eta\sim0.85$, which describes the (cooling) effect of thermal inertia in addition to roughness.
The best-fit $\eta^\prime$ is inversely correlated with best-fit $\Gamma$.
The impact of thermal inertia on surface temperatures is twofold: It changes the response to the eclipse (i.e.\ depth, shape, and wavelength dependence of the eclipse-induced flux drop), but it also determines the diurnal temperature distribution and therefore the overall spectral shape (color temperature).
As will be discussed in the next section, the effective thermal inertia for these two phenomena need not be the same because of the different time and depth scales involved.  Our model, however, assumes a homogeneous thermal inertia.
Due to the design of our observations, the data analysis is more sensitive to the eclipse effect than to the diurnal effect; we take the variation in best-fit $\eta^\prime$ to be an effective description of the residual diurnal effect 
rather than an indication of roughness inhomogeneity.

\section{Discussion}
\label{sect:discussion}

\subsection{Thermal inertia}
\label{sect:discussion:ti}

Our thermal-inertia result 
is below the value for lunar regolith (\unit{50}{\TIunit}), suggesting a very fine regolith and dearth of rocky outcrops.
For a meaningful comparison, however, the lower temperatures at a heliocentric distance of $r\sim\unit{6}{\AU}$ must be taken into account:
For predominantly radiative heat transfer, as expected in a fine regolith on an airless body, the thermal conductivity $\kappa$ scales with $T^3$, hence 
\begin{equation}
\label{eq:scaling}
\Gamma\propto\sqrt{\kappa}\propto T^{3/2} \propto r^{-3/4}.
\end{equation}
The thermal inertia of Patroclus ($20\pm\unit{15}{\TIunit}$ at \unit{5.95}{\AU})
scaled to $r=\unit{1}{\AU}$ is around $\unit{76}{\TIunit}$, comparable to the lunar value and suggesting a similarly fine regolith.

\begin{table}
\caption{Thermal-inertia measurements of Trojans (top 3 lines) and Jovian satellites (below).}
\label{table:TI}
\label{lasttable}
  \centering
  \begin{tabular}{r|lll}
\bottomrule
      & Thermal inertia &Method & Reference \\
      & (\TIunit)       & & \\
\midrule 
(617) Patroclus  & $20\pm15$ & eclipse & this work \\
(2363) Cebriones  & $<14$ & diurnal & \citet{Fernandez2003} \\
(3063) Makhaon    & $<30$ & diurnal & \citet{Fernandez2003} \\
Europa  &45--70 & diurnal & \citet{Spencer1999},  \citet{Greeley2004} \\
Io      &$13\pm4$   & eclipse &     \citet{MorrisonCruikshank1973}  \\
        & $\sim 70$ & diurnal &  \citet{Rathbun2004},  \citet{McEwen2004}\\
Ganymede  & $14\pm2$ & eclipse &     \citet{MorrisonCruikshank1973}  \\
          & $\sim70$ & diurnal & \citet{Spencer1987}, \citet{Pappalardo2004}\\
Callisto  & $11\pm1$ & eclipse &     \citet{MorrisonCruikshank1973}  \\
          & $\sim50$ & diurnal & \citet{Spencer1987}, \citet{Moore2004} \\
\bottomrule    
  \end{tabular}
\end{table}

Our  result is similar to published upper limits on the thermal inertia of two other Trojans \citep[see also \tableref{table:TI}]{Fernandez2003} which were, however, 
based on a significantly less extensive database and a highly indirect method. 
We are not aware of any  other thermal-inertia determination of Trojans.

\Tableref{table:TI} contains a list of 
published thermal-inertia values for Jovian satellites, which are at the same heliocentric distance as Patroclus.
Two values are given for Io, Ganymede, and Callisto, respectively, based on ground-based eclipse observations (first line) or  on thermal imaging 
of the diurnal temperature distribution from near-by spacecraft (second line).
The latter  are effective disk-average values; significant thermal-inertia inhomogeneity was found across the surfaces of the satellites.
Patroclus' thermal inertia is roughly in line with the satellites' eclipse thermal inertia, which is lower than their diurnal thermal inertia by a factor of several.
Different explanations for this difference in the satellites have been published; \citet{Greeley2004} review two mechanisms which require significant amounts of surface ice, and a third mechanism 
\citep[first proposed by][]{MorrisonCruikshank1973} based on
vertical thermal-inertia inhomogeneity:
Because eclipse events are much shorter than the spin period, the corresponding thermal wave penetrates much less deep.  Keeping in mind that the TPM 'thermal inertia' is really an average over the relevant depth scale (see footnote \ref{footnote:depth} on p.\ \pageref{footnote:depth}), 
the effective diurnal thermal inertia is averaged over deeper regions than its eclipse counterpart.
The thermal-inertia difference then implies that near-surface material has a lower thermal inertia than the material underneath (at \cm\ depths), consistent with expectations that the most under-dense and 'fluffiest' material is located on top of more compact material.
Such vertical grain-size sorting has been observed directly in the lunar regolith during the \emph{Apollo} era.

For Patroclus, we do not have a measurement of the diurnal thermal inertia, but the eclipse thermal inertia is similarly low. 
No evidence of surface ice has been found in the spectra of Patroclus \citep{EmeryBrown2003,EmeryBrown2004}, hence
the two ice-based mechanisms to increase the diurnal thermal inertia do not apply. A measurement of the diurnal thermal inertia, combined with our result, would therefore allow  depth-resolved information to be obtained to some extent.

While comparisons with the Galilean satellites are instructive,
we caution that Patroclus is smaller than the satellites by 
more than an order of magnitude.
For an asteroid sample ranging from sub-\km\ near-Earth asteroids to (1) Ceres, \citet{Delbo2007} found a trend whereby larger objects have lower thermal inertia.
\citet{Mueller2006} and \citet{Delbo2009} report constraints on the thermal inertia of $D\sim\unit{100}{\km}$ main-belt asteroids (Lutetia, Lydia, Ekard), similar to Patroclus in size.  While the uncertainties are large, $\Gamma\sim\unit{100}{\TIunit}$ appears to be a typical value in this size range. 
Using \eqref{eq:scaling}, this corresponds to $\Gamma\sim\unit{60}{\TIunit}$ at $r\sim\unit{6}{\AU}$, significantly above our result but consistent with the diurnal thermal inertia of the Galilean satellites.
This is consistent with either a systematically finer regolith on Patroclus (it is unclear what could cause this) or the above-mentioned vertical grain-size sorting.

\subsection{Diameter and mass density}
\label{sect:discussion:drho}

The diameter results from two independent data analyses  (NEATM and  BTPM) agree at the \unit{1}{\%} level and average around
$D_\Area\sim\unit{145}{\km}$ for the area-equivalent diameter of the system as a whole: $D_\Area^2 = D_1^2 + D_2^2$. 
The statistical diameter uncertainty is negligible relative to the systematic uncertainty, although the latter is hard to estimate.
Since it is more realistic, the BTPM would be expected to have lower systematics.
While more work is required to accurately gauge the systematic uncertainty in TPM-derived diameters, 
it was found not to exceed \unit{10}{\%} in the case of near-Earth asteroids \citep[see][]{Mueller2007}. 
Due to its more regular shape and the low phase angle of our observations, Patroclus is less challenging to model than near-Earth asteroids.
We conclude that \unit{10}{\%} is a conservative upper limit on the diameter uncertainty, corresponding to \unit{20}{\%} in albedo (see \eqref{eq:dpvh}). Our final results are: $D_\Area = 145\pm\unit{15}{\km}$ and $\pv=0.045\pm0.009$.

Our result is intermediate between previous estimates:
\citet{SIMPS} obtained $140.9\pm\unit{4.7}{\km}$ (assuming $\eta=0.756$), and \citet{Fernandez2003} obtained $166.0\pm\unit{4.8}{\km}$ (assuming $\eta=0.94$).
The quoted error bars are purely statistical.  We point out that their results were subject to significant systematic uncertainties due to the assumed (not fitted to the data) $\eta$ values.
Systematic uncertainties are acknowledged by \citeauthor{Fernandez2003}\ to limit the accuracy of their results, but no quantitative discussion is given. Since our database has a significantly larger wavelength coverage and since we use a more realistic thermal model, 
the true uncertainty in the cited  diameters should be large compared to that in  our results.
With this in mind, all diameter results are in reasonable agreement with one another.

Assuming a diameter ratio of $D_1/D_2=1.082$,
the components' diameters are $D_1=106\pm11$ and $D_2=98\pm\unit{10}{\km}$.
The corresponding bulk density is
$1.08\pm\unit{0.33}{\gramm\usk\power{\cm}{-3}}$,
in excellent agreement with the \citet{PatroNew} result of 
$\unit{$0.90^{+0.19}_{-0.33}$}{\gramm\usk\power{\cm}{-3}}$
and  indicative of a bulk composition  dominated by water ice. Significant contributions by heavier materials (such as rock) are possible if compensated by interior voids.
Near-infrared spectra show no indication of water ice on the surface, even in the \unit{3}{\micron} region, where even very small amounts of \water\ would be detected \citep{EmeryBrown2003,EmeryBrown2004}. If the bulk composition does include significant \water, the surface must be coated by a devolatilized mantle.
This would be expected for a extinct-comet-like object, i.e.\ an icy object that spent significant amounts of time in the inner Solar System.

It is not unusual for published asteroid diameters to be quoted with purely statistical uncertainties, neglecting the typically dominant systematics. 
Most known asteroid diameters are derived from thermal-infrared observations.
While more research is needed to quantify their systematic diameter uncertainty, \unit{10}{\%} seems to be an appropriate value, 
larger when the color temperature is assumed rather than fitted to the data, 
as is typically done in the widely used 'Standard Thermal Model' (STM).
This includes the entire SIMPS catalog \citep{SIMPS}, i.e.\ the majority of currently known asteroid diameters.

A realistic evaluation of the size uncertainty---including systematics---is crucial when published diameters are used to determine the mass density.
Due to the dependence on $D^{-3}$, density uncertainties are frequently dominated by that in diameter. Ignoring the systematic diameter uncertainty may lead to overly optimistic density estimates.
A diameter uncertainty of \unit{10}{\%} translates into \unit{30}{\%} in density; \unit{15}{\%} in diameter means some \unit{45}{\%} in density.

\subsection{Mineralogy}
\label{sect:discussion:mineralogy}

It is instructive to compare Patroclus to its mythological antagonist, (624) Hektor, the only other currently known Trojan binary and the only other Trojan with known mass density.

Patroclus and Hektor have nearly identical albedo and mid-IR emissivity spectra, yet their densities differ by a factor $>2$: $\rho\sim2.2$--\unit{2.5}{\gramm\per\centi\metre\cubed} for Hektor \citep[e.g.][]{Weidenschilling1980,Marchis2006b,Lacerda2007} to Patroclus' $\rho\sim\unit{1}{\gramm\per\centi\metre\cubed}$.
Reflectance spectra of both are featureless from $\sim0.3$ to \unit{4.0}{\micron}, but Hektor displays a much steeper spectral slope throughout that wavelength range. It is uncertain whether the difference in spectral slope indicates distinct surface compositions or is a manifestation of differences in, e.g., grain size, abundance of opaque materials, or degree of space weathering.

Trojans in general form two distinct groups based on visible and near-IR colors and slopes \citep{Szabo2007,Emery2007,Melita2008,Roig2008}, one group similar to Hektor and one similar to Patroclus.
It is tempting to take the density difference between Patroclus and Hektor as indicating a compositional difference between the two groups. 
In the light of this, their very similar mid-IR spectra are intriguing.

The mutual orbits of the Patroclus and Hektor systems are conspicuously dissimilar: 
While Patroclus is a double-asteroid system with two components of roughly equal size and nearly spherical shape, 
Hektor consists of a large irregular primary (which is thought to be a contact binary itself) and a much smaller moonlet \citep{Marchis2006b}.

\section{Conclusions}
\label{sect:conclusions}

\paragraph{Patroclus}
\label{sect:conclusions:Patroclus}

Patroclus has a low thermal inertia of $20\pm\unit{15}{\TIunit}$. This first thermal-inertia measurement for a Trojan indicates a surface covered in fine, mature regolith similar to large main-belt asteroids and the Galilean satellites.
Independent evidence for a fine regolith is provided by the mid-IR emissivity features reported herein, which imply the presence of very fine silicate grains on the surface.

Patroclus' optical lightcurve implies a nearly spherical shape for both components and indicates that the system is fully synchronized, probably due to tidal component-component interactions.

We confirm that Patroclus' albedo is low ($\pv=0.045\pm0.009$) and that the system's 
bulk mass density is $1.08\pm\unit{0.33}{\gramm\usk\power{\cm}{-3}}$.
Patroclus' known physical properties (albedo, density, fine silicate regolith, and published optical spectra which are similar to cometary surfaces but devoid of any traces of water ice)
match those expected in extinct comets and suggest that Patroclus formed beyond the 'snow line' but spent significant amounts of time in the inner Solar System, leaving its surface devolatilized.
Such a dynamical history is expected for Trojans in the framework of the 'Nice Model' \citep{Morbidelli2005,Morbidelli2009}.
While Patroclus is arguably the best-studied Trojan asteroid, it is unclear how representative it is for the Trojan population as a whole; the composition of the only other Trojan with known mass density, Hektor, is dominated by  materials heavier than water ice.

\paragraph{Binary eclipses in the thermal infrared}
\label{sect:conclusions:method}

We demonstrate the viability of a new method to measure the thermal inertia of asteroids, through thermal observations during and after shadowing events in binary systems.
 This new method is considerably more direct than the usual method of sampling the diurnal temperature distribution through observations at a large phase-angle range.

In either case, the measured thermal inertia is an effective average over the relevant depth scale, which is smaller for eclipses than for rotation due to the shorter time scales involved.
Vertical  grain-size sorting may cause the 
eclipse thermal inertia to be lower than the diurnal thermal inertia.
While this  makes comparisons harder, it may shed light on the elusive vertical regolith structure.

For objects in the outer Solar System, 
sufficient phase-angle coverage for the diurnal measurement method
 is hard or impossible to obtain.
The maximum phase angle $\alpha_\text{max}$ decreases with increasing
 heliocentric distance $r$ (in \AU):
$\alpha_\text{max}=\arctan(1/r)$. For Neptune, e.g., 
$\alpha_\text{max}\sim\unit{2}{\degree}$. 
Thermal eclipse observations may be the only  method to measure the thermal inertia of  Centaurs and Kuiper Belt objects from (close to) Earth --
once the mutual orbits of such binary systems are reliably known.

\section{Acknowledgments}
We wish to thank Petr Pravec for helping develop this project.
Enlightening discussions with John Spencer (and a great mountain hike!) are gratefully acknowledged.
This work is based on observations made with the Spitzer Space Telescope, which is operated by the Jet Propulsion Laboratory, California Institute of Technology under a contract with NASA. Support for this work was provided by NASA through an award issued by JPL/Caltech.
The work of M.M.\ is supported through JPL/Spitzer contract \#1287372.
The work of F.M.\ is supported by the National Aeronautics and Space Administration issue through the  Science Mission Directorate Research and Analysis Programs number NNX07AP70G \& NNG05GF09G




\label{lastpage}

\end{document}